\documentclass[letterpaper, paper,twoside,notoc]{JHEP}
\def\sm{Standard Model}
\def\gesim{\lower0.5ex\hbox{$\:\buildrel >\over\sim\:$}}
\def\lesim{\lower0.5ex\hbox{$\:\buildrel <\over\sim\:$}}

\def\sqr#1#2{{\vcenter{\hrule height.#2pt
      \hbox{\vrule width.#2pt height#1pt \kern#1pt
         \vrule width.#2pt}
      \hrule height.#2pt}}}
\def\square{{\mathchoice\sqr56\sqr56\sqr{2.1}3\sqr{1.5}3}}
\def\inv#1{{1\over#1}}

\def\ocal{{\cal O}}
\def\lcal{{\cal L}}
\def\ie{{\it i.e.}}
\def\eg{{\it e.g.}}
\def\lr{{\lcal_\ell}}
\def\lh{{\lcal_h}}
\def\leff{{\lcal_{eff}}}
\def\lredu{{\lcal_{red}}}
\def\wt{\widetilde}
\def\ti{{\tilde{j}}}

\title{Effective $\beta$-functions for Effective Field Theory}
\author{Martin B. Einhorn\\ Michigan Center for Theoretical Physics, \\University of Michigan,
Ann Arbor, Michigan, 48019-1120}
\author{J. Wudka\\Department of Physics, \\ University of California, Riverside CA 92521-0413.}
\abstract{We consider the problem of determining the beta-functions for any reduced effective field theory.  Even though not all the Green's functions of a reduced effective field theory are renormalizable, unlike the full effective field theory, certain effective beta-functions for the reduced set of couplings may be calculated without having to introduce vertices in the Feynman rules for redundant operators. These effective beta-functions suffice to apply the renormalization group equation to any transition amplitude (\ie, S-matrix element), thereby rendering reduced effective field theories no more cumbersome than traditionally renormalizable field theories.  These
effective beta-functions may equally be regarded as the running of 
couplings for a particular redefinition of the fields.}


\begin{document}

\section{Introduction}

All local field theories describing natural phenomena are probably effective field theories, describing relevant physics over a limited range of energies or mass scales.  Even though they contain interactions that are associated with both renormalizable and nonrenormalizable vertices, gradually we have come to appreciate that they should be regarded as perturbatively renormalizable.\cite{weinberg}  Although this requires the presence of an infinity of vertices, to any given degree of experimental accuracy, they may be approximated by a finite number of vertices.  Thus, like traditional renormalizable field theories, they can describe observable phenomena in terms of a finite number of parameters.  An effective field theory  breaks down at an energy scale at which new physics enters, usually because some of the vertices previously regarded as local become nonlocal owing to a change in the fundamental degrees of freedom.  Typically, in particle physics, this corresponds to a threshold for the production of new particles.  

Of course, as higher dimensional operators are introduced, the multiplicity of vertices grows rapidly, so there is a proliferation of coupling constants to be fit to data.  The situation is made simpler than it might otherwise be by the observation that many higher-order operators are redundant, in that S-matrix elements only depend on a subset of their associated couplings.  The redundancy is expressed by a basic theorem\cite{equivthm,georgi,arzt} about ``equivalent field theories."  This theorem takes several different forms, the one that will be most useful to us is the form expressed in Ref.~\cite{arzt}, viz., two field theories have the same S-matrix elements if the difference of their Lagrangians vanishes upon application of the ``classical" equations of motion.\footnote{This is equally true whether expressed in terms of bare fields and couplings or renormalized fields and couplings.  This redundancy is distinct from the statement that two Lagrangians are equivalent if their difference is a total divergence.  This latter equivalence, which perturbatively depends only on conservation of momentum, leaves Green's functions, rather than only S-matrix elements, unchanged and will be regarded as trivial in the present context.}  When the higher dimensional vertices are to be used only in tree approximation, this statement is sufficient.  One arbitrarily chooses any convenient, linearly-independent subset, and parameterizes S-matrix elements accordingly.  The Lagrangian associated with a particular linearly-independent subset will be referred to as the ``reduced effective field theory."   

When higher-dimensional operators are used in loop corrections, new issues arise.  In general, because of operator mixing, one will encounter divergences of the form of some of the redundant operators, \ie, the Green's functions are not renormalizable if one does not include counterterms for operators that have been eliminated.\cite{arzt}  On the other hand, useful information about the theory, such as its beta-functions and anomalous dimensions, must be determined from an analysis of the Green's functions.  To calculate radiative corrections then,  it would seem necessary to include all vertices and not simply a linearly-independent subset.  

We will show in the following that, for applications of the renormalization group to observables, it is unnecessary to include redundant vertices in the Feynman rules for the effective field theory.  Even though counterterms may be required that involve redundant operators, one may reinterpret them in terms of counterterms for the initially chosen, linearly-independent set and thereby determine the running of these couplings directly.   Thus, equipped with a dictionary for re-expressing each redundant operator in terms of linear combinations of the reduced set, one may determine the \emph{effective beta-functions} for all the couplings of the reduced effective field theory.  

Running couplings for renormalizable vertices are by now quite familiar.  To cite two well-known examples, we refer to the renormalization of the electromagnetic coupling constant from the scale of the electron mass up to the electroweak scale\cite{sirlin} or to the running of the \sm\ gauge and Yukawa couplings from the electroweak scale up to the unification scale.\cite{running} The running of couplings for higher dimensional operators is somewhat less familiar but can be important nevertheless for relating phenomena over a wide range of scales,\eg, matching the parameters of the Standard Model at the electroweak scale to certain precision electroweak measurements at low energies, such as $g_\mu\!-\!2$,\cite{Arzt:1994wz} rare K-decays\cite{Littenberg:1993qv}or B-decays,\cite{Simma} parity violation in atoms\cite{Bouchiat:1980gp} or nuclei,\cite{Holstein:1988ei} {\it etc.}\cite{Langacker}   Similar issues arise if one simply wants to use higher dimensional operators to parameterize new physics without prejudice about the underlying theory.\cite{patterns}  Moreover, if one is ever successful in relating the physics of quantum gravity or M-theory to laboratory experiments, it most likely will be by means of an effective field theory description over widely disparate energy scales.  Supergravity theories are prime examples of such nonrenormalizable effective field theories.  Thus, we anticipate that the results obtained in this paper will be of widespread applicability and hope that the associated calculational simplifications will be of substantial utility.

\section{Reduced Coupling Constants and Effective $\beta$-functions}

To pose the issue precisely, suppose we are given an effective field theory of generic form 
\begin{equation}
\leff=\lr + \lh
\label{leff}\end{equation}
where $\lr$, often referred to as the ``classical" Lagrangian, typically contains the free-field equations and possibly certain other low-order monomials, and $\lh$ represents the infinity of {\bf all} possible higher-order operators that are consistent with the presumed symmetries of the fundamental theory.  In what we shall term the \emph{traditional} perturbative applications, $\lr$ consists of the infrared relevant and marginal operators, the so-called ``renormalizable" vertices, and $\lh$ contains all infrared irrelevant operators, traditionally called the ``non-renormalizable" vertices.  However, other splits can be and sometimes have been made.  For example, in chiral perturbation theory, the expansion in ``chiral-dimension" differs from the usual expansion in engineering dimensions.  For expansions other than the traditional perturbation series, such as $1/N$-expansions, the split would be determined by the order of   vertices in powers of $1/N.$  Even within perturbation theory, there is arbitrariness.  One might choose $\lr$ to consist of the free or superrenormalizable theory and $\lh$ to consist of all remaining vertices, a procedure essentially equivalent to one described by Georgi.\cite{georgi}  In gauge theories, however, it is generally advantageous to make a gauge-invariant split, so that both $\lr$ and $\lh$ are each gauge-invariant.\footnote{Although we will usually have in mind applications in four space-time dimensions, our discussion is valid in any number of dimensions and for Minkowski or Euclidean signature.}

So long as $\lh$ involves some small parameter that enables its interactions to be treated perturbatively, this formalism can be used.  It is important that $\lr$ and $\lh$ include all operators consistent with the presumed symmetries.  The masses and couplings appearing in $\lr$ will be denoted as  $\lambda_a$; those appearing in $\lh$, as $\alpha_i$.  We therefore write $\lh$ as
\begin{equation}
\lh=\sum_i \alpha_i\ocal_i. 
\label{higherorder}\end{equation}
In the traditional situation in four dimensions, the operators $\ocal_i$ are all the (gauge-invariant) monomials of dimension five and higher formed from the fields and their covariant derivatives.  Correspondingly, the associated coupling constants $\alpha_i$ have dimensions of an inverse mass, say $\Lambda,$ to some power.   We expect the effective field theory to describe phenomena for some range of momenta $p^\mu\!\ll\Lambda.$\footnote{In reality, there may be several scales associated with the onset of new physics, but one may think of $\Lambda$ as the smallest of them, in which case, some of the couplings $\alpha_i$ may appear to be extremely small, tacitly involving the ratio of very different physical energy scales.}  If, besides local gauge symmetries, $\lr$ has global symmetries, then one may need to discuss whether they are \emph{accidental,}\cite{weinberg2} in which case $\lh$ need not respect them, or \emph{natural,}\cite{thooft} in which case, there may be implicit restrictions on the coefficients of the higher dimensional operators.  For example, if $\lr$ acquired some chiral symmetry in the limit that some mass or Yukawa coupling vanished, then in any \emph{natural} theory, we would generally expect the couplings $\alpha_i$ in $\lh$ to respect that symmetry as well.

As remarked in our introduction, the monomials $\{\ocal_i\}$ form an over-complete set.  By application of the equations of motion associated with $\lr$, one may replace them by a subset $\{\wt{\ocal_\ti}\}$ of linearly-independent operators,\cite{arzt} thereby substantially reducing the number of parameters $\alpha_i$.\footnote{Henceforth, generic indices are denoted by $i,j,\ldots,n$; indices of the linearly-independent subset, by $\tilde{i},\tilde{j},\ldots,\tilde{n}$; and indices of redundant operators or couplings, by $p,q,\ldots.$}  Even if one were not already aware of the equivalence theorem, one would discover that observables, \ie, S-matrix elements, depended only a reduced set of couplings which may be written in the form 
\begin{equation} 
\wt{\lambda_a}=\lambda_a + A_{ap}(\lambda_a)\alpha_p,\qquad
\wt{\alpha_\ti}=\alpha_\ti+B_{\ti p}(\lambda_a)\alpha_p. 
\label{alpharedu}\end{equation}
The sum over redundant couplings $\alpha_p$ in Eq.~(\ref{alpharedu}) may in general extend over an infinite number of higher-dimensional couplings, because the product of a power of a mass or superrenormalizable coupling from $\lr$ with a nonrenormalizable coupling $\alpha_p$ will have dimension of some lower-order $\alpha_i$ or even $\lambda_a.$\footnote{Ordinarily, the possibility that interactions in $\lh$ might modify the couplings in $\lr$ is ignored, an approximation that is justified only to the extent that the ratio of masses in $\lr$ to mass scales $\Lambda$ characteristic of $\lh$ are small compared to the appropriate renormalized couplings.  This is a naturalness constraint usually tacitly assumed.}
In this way, one arrives at a reduced effective Lagrangian\cite{arzt} 
\begin{eqnarray} 
\wt{\leff}&=&\wt{\lr}+\wt{\lredu},\\
{\rm where~}\wt{\lredu}&\equiv& \sum \wt{\alpha_\ti} \wt{\ocal_\ti},
\label{lredu}\end{eqnarray}
and $\wt{\lr}$ has the same form as the original $\lr$ with the $\lambda_a$ replaced by the $\wt{\lambda_a}.$  The linearly-independent set $\{\wt{\ocal_\ti}\}$ is presumed to be complete, in the sense that any redundant operator $\ocal_i$ may be written as a linear combination of the operators appearing in $\lredu.$

It may clarify the preceding language by reference to a common example.  Suppose $\lr$ is the theory of a real scalar field
\begin{equation}
\lr=\frac{1}{2}(\partial_\mu \phi)^2 -\frac{m^2}{2}\phi^2 - \frac{\lambda}{4!}\phi^4
\end{equation}
The equations of motion that follow from this are
\begin{equation}
-\partial^2\phi=m^2\phi+ \frac{\lambda}{3!}\phi^3
\end{equation}
Because of the discrete symmetry, $\phi\to-\phi$, there are no dimension-five operators contributing to $\lh.$
There are three dimension-six operators (up to integration-by-parts)
\begin{equation}
\lh= -\frac{\alpha_1}{6!}\phi^6 - 
\frac{\alpha_2}{4} \phi^3(\partial^2 \phi) -
\frac{\alpha_3}{2} (\partial^2\phi)^2
\end{equation}
These three are not linearly independent.  For example, one may replace the third operator $ (\partial^2\phi)^2$ with its equivalent $(m^2\phi+ {\lambda}\phi^3/{3!})^2$ yielding a linear combination of operators from $\lr$ and $\lh,$ whose specific form is not needed.  One may proceed similarly also with the second operator.  The important thing to note is that the original $\leff$ has been replaced by a reduced $\lredu$ in which the number of operators of dimension-six has diminished from three to one.  Of course, as one proceeds to consider even higher dimensional operators or more complicated theories involving several different fields, the number of redundant operators will rapidly increase.

Returning to the general case, although there is no problem of principle involved in working with the full $\leff,$  it can be extremely cumbersome to carry along vertices and couplings that, when observables are calculated, are redundant.  For example, in the \sm, there are 81 linearly-independent dimension-six operators\cite{buchwyl} and who knows how many redundant ones.  Yet, in applications, one frequently wants to employ the renormalization group to compare parameters at very different scales and to reduce the dependence of perturbation theory on large logarithms.  This can be of use even in the \sm\ when comparing predictions for a low-energy measurement with the parameters determined at the weak scale.  In such cases, given the parameters associated with the scale of vector boson masses $M_W$ and $M_Z$, one may wish to determine an effective Lagrangian at scales approximating $\Lambda_{QCD}$.  This $\leff$ would include the renormalizable terms associated with QED and QCD as well as higher dimensional operators contributing to observables, such as a five-dimensional, Pauli-type interaction contributing to $g\!-\!2$ or a six-dimensional,  four-fermion interaction responsible for beta-decay.  Moreover, the scale dependence of these higher-dimensional couplings $\alpha_i,$ since they correspond to summing leading logs, can sometimes be used to estimate the most important  contributions to the loop corrections in next higher order.\cite{dgg}  

In general, the beta-functions for $\leff$ are functions of $\lambda_a,\,\alpha_i$.  In a mass-indepen\-dent renormalization prescription, such as $\overline{MS},$ these take the form
\begin{equation} 
\mu\frac{\partial\lambda_a}{\partial\mu} = \beta_a(\lambda_b,\alpha_i)\qquad
\mu\frac{\partial\alpha_i}{\partial\mu} = \beta_i(\lambda_b,\alpha_i).
\label{betafns}\end{equation}
As remarked in the last footnote, we emphasize again that the possible dependence of $\beta_a$ on the couplings $\alpha_i$ that is generally ignored should, in the present context, be taken into account.  Is there any way to simplify calculations so that the Feynman rules\footnote{By ``the Feynman rules," we have in mind the by-now standard formulation in terms of renormalized fields and couplings plus appropriate counterterms rather than in terms of bare fields and couplings.} involve only the reduced set of vertices $\{\wt{\ocal_\ti}\}$?  Since not all Green's functions for the reduced theory can be renormalized, one would anticipate that the answer is no.  Remarkably, we \emph{can} determine the running of the reduced set of couplings $\wt{\alpha_\ti}$ without reintroducing the redundant couplings into the Feynman rules.  The trick is to reexpress the counterterms for the redundant operators back in terms of the linearly-independent set, just as one used in going from $\leff$ to $\lredu.$ 

To see this without having to perform a detailed analysis order-by-order in perturbation theory, we recall that an arbitrary S-matrix element $S$ of a theory satisfies the homogeneous renormalization group equation
\begin{equation}
\left[\mu\frac{\partial}{\partial\mu} + 
\beta_a\frac{\partial}{\partial{\lambda_a}} + 
\beta_i\frac{\partial}{\partial{\alpha_i}}\right]S = 0.
\label{rge}\end{equation}
However, because of the equivalence theorem,\cite{equivthm,arzt} the S-matrix depends only on the  reduced set of couplings defined in Eq.~(\ref{alpharedu})
\begin{equation}
S=S\left(\wt{\lambda_a}, \wt{\alpha_\ti}\right).
\label{smatrix}\end{equation}
Given the relation Eq.~(\ref{alpharedu}) between the reduced couplings and the full set, we may use the chain rule to rewrite Eq.~\ref{rge} as
\begin{equation}
\left[\wt{\beta_a}\frac{\partial S}{\partial{\wt{\lambda_a}}} +
 + \wt{\beta_\ti}\frac{\partial S}{\partial\wt{\alpha_\ti}} 
\right] = -\mu\frac{\partial S}{\partial\mu},
\label{rgetilde}\end{equation}
where the \emph{effective beta-functions} are defined by 
\begin{equation}
\wt{\beta_a}\equiv \beta_a+A_{ap}\beta_p + 
\beta_b\frac{\partial A_{ap}}{\partial\lambda_b}\alpha_p,\qquad
\wt{\beta_\ti}\equiv \beta_\ti + B_{\ti p}\beta_p + 
\beta_a\frac{\partial B_{\ti p}}{\partial \lambda_a}\alpha_p.
\label{betatilde}\end{equation}
It is clearly these effective beta-functions that govern the scale dependence of S-matrix elements, so it is these that we would like to extract from the effective field theory.  It would be economical if we could work from the reduced effective field theory $\lredu$ rather than having to restore redundant couplings to form the combinations in Eq.~(\ref{betatilde}).  
This can be done! 

\section{The Effective $\beta$-Function Theorem}

We will now establish the following theorem:  the effective beta-functions, despite their complicated definitions in Eq.~(\ref{betatilde}) involving the full, renormalizable $\leff$,  are functions of the reduced set of coupling constants of $\lredu.$  The first step to establish this is to note that each of the derivatives of the S-matrix in Eq.~({\ref{rgetilde}) is a function only of the reduced couplings $\wt{\lambda_a}, \wt{\alpha_i}.$  We would like to infer that their coefficients, $\wt{\beta_a}, \wt{\beta_\ti}$, also depend only on the reduced couplings, \ie,  that
\begin{equation}
\wt{\beta_a}=\wt{\beta_a}(\wt{\lambda_a}, \wt{\alpha_{\tilde{j}}}),\qquad
\wt{\beta_\ti}=\wt{\beta_\ti}(\wt{\lambda_a}, \wt{\alpha_{\tilde{k}}}).
\label{betaeff}\end{equation}
To establish this, we may think of the $\wt{\beta_a}, \wt{\beta_\ti}$ in Eq.~(\ref{rgetilde}) as ``unknowns" to be solved for and the various derivatives of the S-matrix as ``knowns."  For the sake of argument, we should (temporarily) terminate the order of the couplings at some finite number.\footnote{Although such a finite number is not strictly speaking renormalizable, we may in principle include as many terms as necessary to achieve any desired level of accuracy.}  Since the unknowns $\wt{\beta_a}, \wt{\beta_\ti}$ are in one-to-one correspondence with the reduced couplings $\wt{\lambda_a},\wt{\alpha_\ti},$ we need as many independent equations as reduced couplings in order to contemplate solving the system.  We can generate as many equations as we need by choosing different S-matrix elements or even by choosing different values for the momenta for the same S-matrix element (whose dependence on momenta has been suppressed above). So long as the system is nonsingular, we may invert and express these $\wt{\beta_a}, \wt{\beta_\ti}$ in terms of quantities manifestly depending only on the reduced set. Having established this result for an arbitrary finite number of terms, we may then extend the sums in Eq.~(\ref{betaeff}) to infinity.  Thus, the effective beta-functions, $\wt{\beta_a},\wt{\beta_\ti}$, are indeed functions of the reduced set of couplings $(\wt{\lambda_a},\wt{\alpha_\ti})$ only!  This result is one of the main conclusions of this paper from which numerous consequences flow.

Suppose we set all the redundant couplings $\alpha_p$ to zero in the defining equation for the effective beta-functions Eq.~({\ref{betatilde}).  According to Eq.~(\ref{alpharedu}), then $\wt{\lambda_a}={\lambda_a}$ and $\wt{\alpha_\ti}=\alpha_\ti,$  so that our result Eq.~(\ref{betaeff}) reduces to 
\begin{equation}
\wt{\beta_a}=\wt{\beta_a}(\lambda_a, \alpha_\ti)=\beta_a+A_{ap}(\lambda_b)\,\beta_p,\qquad
\wt{\beta_\ti}=\wt{\beta_\ti }(\lambda_a, \alpha_\ti)=\beta_{\tilde{k}} + B_{\ti p}(\lambda_a)\,\beta_p.
\label{betaefftilde}\end{equation}
The only couplings $\alpha_\ti$ that appear in arguments of all the beta-functions are those associated with the linearly-independent subset appearing in the reduced effective Lagrangian Eq.~(\ref{lredu}).  Even though we have reduced the number of vertices to those of $\lredu,$ it nevertheless still appears to be necessary to calculate the $\beta_p$ for the redundant operators as well as the individual $\beta_a, \beta_i$.  Notice, however, that \emph{the linear combinations of the beta-functions $\beta_a, \beta_i$ appearing in Eq.~(\ref{betaefftilde}) are identical to those appearing in the definition of the reduced coupling constants in Eq.~({\ref{alpharedu}).}}}  Therefore, we may finesse the calculation of the individual components $\beta_a,$ $\beta_i,$ and $\beta_p,$  and calculate the effective beta-functions $\!  directly.$  The procedure may be described as follows:  In the Feynman rules, one need only include vertices associated with the reduced effective Lagrangian Eq.~({\ref{lredu}).  The counterterms required to renormalize various Green's functions will involve local operators $\ocal_p$ that do not appear in the reduced effective Lagrangian $\lredu.$  However, in order to determine the effective beta-functions, one may simply apply the equations of motion to reexpress those counterterms in terms of the reduced set of operators and then to calculate the $\wt{\beta_a},\wt{\beta_\ti}$ in the standard fashion.  

In short, one may treat counterterms for the complete set $\lh$ as if they were counterterms for the reduced set $\lredu,$ with the effective beta-functions calculated as if the operators were truly equivalent.  Thus, equipped with a translation guide to operator equivalences, the calculations for the reduced effective theory becomes no more cumbersome than for a traditionally renormalizable field theory.  For practical purposes, the nonrenormalizability of Green's functions based on the vertices of $\lredu$ alone is irrelevant.

One should not be misled by the linear relation between the reduced couplings and the original set in Eq.~(\ref{alpharedu}).  Our result does not require that one work only to first-order in $\lh,$  even though, 
 in most applications, it is usually sufficient to do so.  Nowhere does the proof assume that our effective beta-functions must be linear in $\alpha_\ti.$

The occurrence of an infinite number of $\alpha_p$ in Eq.~(\ref{alpharedu}) and of $\beta_p$ in Eq.~(\ref{betaefftilde}) results from the presence of masses and superrenormalizable couplings in $\lr.$  As remarked earlier, often times their effects on the beta-functions may be neglected  if their ratios to the scales in $\lh$ are sufficiently small.  As a corollary of our result, suppose that such mass ratios can be ignored.  Then, in a mass-independent renormalization prescription, we may conclude that $A_{ap}=0$ and that $\wt{\beta_a}=\beta_a(\lambda_a),$ as usual for the ``renormalizable" couplings.  Moreover, the only non-zero mixings $B_{\ti p}$ that occur in Eq.~(\ref{alpharedu}) involve only those finite number of couplings $\alpha_p$ having the same dimension as $\alpha_\ti$.  Therefore, the effective beta-functions for the ``non-renormalizable" couplings involve only a finite number of terms in Eq.~(\ref{betaefftilde}).  This is the form in which they appear in most applications to date.

Although we have derived our effective beta-functions Eq.~(\ref{betaefftilde}) by reference to the S-matrix, there is another sense in which they represent the beta-functions of the field theory associated with given reduced set of operators Eq.~(\ref{lredu}).  Inasmuch as we may eliminate redundant operators by a field redefinition, we may regard the renormalization procedure as determining not only the counterterms but also the field appropriate to a given basis set.  That is, at each order in the perturbation expansion, we may make a field redefinition to remove any counterterms for redundant operators, as described below Eq.~(\ref{betaefftilde}).  In this way, one may regard the theory as renormalizable within the particular linearly independent, complete basis set $\{\wt{\ocal_\ti}\}$ by appropriate choice of the field.  With this natural prescription, our effective beta-functions may be regarded as the beta-functions for the reduced field theory itself.  From this point of view, the S-matrix is simply a crutch enabling us to show that this convention is precisely what is required for physical applications and demonstrating how the effective beta-functions would be related to the beta-functions associated with other possible field definitions.

\section{Example:  $(\phi^3)_6$}

To illustrate the general discussion above and verify the theorem for a special case, we consider here the simple example of $\phi^3$-theory in six dimensions.\footnote{Of course, lacking a sensible ground state in the infrared regime, $(\phi^3)_6$ only exists formally as a perturbation expansion.  Nevertheless, it is perhaps the simplest nontrivial example at one-loop order.} 
The effective Lagrangian (including terms of dimension eight) is ${\cal L}=\lr+\lh,$ with
\begin{eqnarray}
\lr& \equiv& \inv2 \left( \partial \phi \right)^2 - \inv2 m^2 \phi^2 - \sigma \phi - \inv{3!} f \phi^3\cr
\lh&\equiv& -{\alpha_1 \over 4!\Lambda^2 } \phi^4  +  
{\alpha_2 \over 2 \Lambda^2 } \phi^2 \square \phi  
- {\alpha_3 \over 2 \Lambda^2 } \left( \square \phi \right)^2 + \ldots
\label{lphi3}\end{eqnarray}
Since for these purposes, we will only be interested in the role of the dimension-eight operators, we have changed our notation slightly, replacing $\alpha_i$ by $\alpha_i/\Lambda^2,$ where $\Lambda$ represents the scale of new physics.  As is customary, we will assume that, at each order in the loop expansion,  the field is shifted by an appropriate constant so as to eliminate the linear term, effectively setting $ \sigma =0. $   The calculation of the one-loop beta-functions (in the $\overline{MS}$ scheme) is straightforward  (however tedious).  Keeping only terms linear in $ \alpha_i $ and defining $ \eta \equiv (m /\Lambda)^2,$ we find 
\begin{eqnarray}
\beta_\eta &=&  \eta \left[- \inv3 f^2 +
   \eta \left( \inv2 \alpha_1 + {16\over3} f \alpha_2 +{10\over3} f^2
\alpha_3 \right)  \right] \cr 
\beta_f &=& -{3\over4} f^3 - \eta \left( {5 \over2} f \alpha_1 + {37\over2}
f^2 \alpha_2 + 11 f^2 \alpha_3 \right) \cr
\beta_{\alpha_1} &=& -\left[{17\over3} f^2 \alpha_1 + 24 f^3 \alpha_2 + 12 f^4
\alpha_3\right] \cr
\beta_{\alpha_2} &=&  \inv6 f \alpha_1 - \inv{12} f^2 \alpha_2 + \inv3
f^3 \alpha_3\cr
\beta_{\alpha_3} &=&\,  \inv3 f \alpha_2 +\, \inv6 f^2 \alpha_3
\label{betas}\end{eqnarray}
The ``classical" equation of motion associated with $\lr$ is \vspace{-.1in}
\begin{equation}
-\square \phi = m^2 \phi + \inv2 f \phi^2,
\vspace{-.1in}
\end{equation}
which we use in $\lh$ to eliminate the dimension-eight operators proportional to $
\alpha_{2}$ and $\alpha_{3}.$ Then the reduced Lagrangian becomes simply 
\begin{eqnarray}
\wt{{\lredu}}&=&  \inv2 \left( \partial \phi \right)^2 - \inv2 \wt{m^2}
\phi^2 - \inv{3!} \wt{f} \phi^3 - {{\wt{\alpha_1}} \over 4!\Lambda^2  } \phi^4,\label{lphi3redu}\\
{\rm where}~~\wt{m^2} &=& m^2 (1 + \eta \alpha_3)\cr
\wt{f} &=& f + 3 \eta ( \alpha_2 + \alpha_3 f)\cr
\wt{\alpha_1} &=&  \alpha_1+6 f  \alpha_2+3 f^2  \alpha_3.
\label{phi3alpharedu}\end{eqnarray}
Using relations (\ref{phi3alpharedu}) and (\ref{betas}), we find by explicit evaluation of Eq.~(\ref{betatilde})
\begin{eqnarray}
\wt{\beta_{\wt{\eta}}} &=& - \wt{\eta}\left[\inv3 \wt{f^2} + \inv2 \wt{\eta} \wt{\alpha_1}\right] \cr
\wt{\beta_{\wt{f}}} &=& -{3\over4} \wt{f^3} - 2 \wt{\eta} \wt{f} \wt{\alpha_1}\cr
\wt{\beta_{\wt{\alpha_1}}} &=& -{14\over3} \wt{f^2}  \wt{\alpha_1}
\label{effbetas}\end{eqnarray}
where $ \wt{\eta} \equiv  \wt{m^2}/\Lambda^2 $. 
Thus, these expressions depend only on the reduced quantities, as guaranteed by our general theorem.   Further, they may be calculated directly from the Feynman rules for the 
reduced Lagrangian Eq.~(\ref{lphi3redu}), thereby simplifying the calculation enormously.

Notice that the ratio of scales $\wt{\eta}$ acts like another dimensionless coupling constant affecting the running of $\wt{f}$.  Even though one is below the threshold of $O(\Lambda)$ for the production of particles associated with new physics, these expressions show clearly that it is relative size of $\wt{\eta}\wt{\alpha_1}$ to $\wt{f}^2$ that determines whether such mass ratios can be neglected.

Let us reflect on our previous remarks concerning the considerable arbitrariness of the split between $\lr$ and $\lh$.  Had we included the $f$ coupling in $\lh$, the intermediate steps would have been somewhat simpler, but the final result Eq.~({\ref{effbetas}) would have been identical.  Aside from verifying the theorem, all we really care about in practice is the result.

\section{Summary \& Discussion}
We have shown that, even though we use a reduced effective Lagrangian and a correspondingly reduced set of vertices in an effective field theory, we may nevertheless define effective beta-functions that determine the running of the associated renormalized coupling constants.  This can be accomplished without having to introduce redundant operators in order to renormalize all the Green's functions.  The scale dependence of all the couplings entering observables may therefore be determined from Feynman rules involving only the reduced set of vertices.  

From the point of view of effective field theory, it is more natural to talk about the running and mixing of coupling constants rather than to refer to the anomalous dimensions and mixing matrix for composite operators.   Although language is largely a matter of taste, by so doing, all vertices are treated on an equal footing. Moreover, knowing the anomalous dimension matrix  is equivalent to the effective beta-functions only when the higher-dimensional operators are treated linearly. 

So long as a complete set of linearly-independent operators has been chosen for the reduced Lagrangian, the results for observables {\bf will} be independent of the choice of reduced operators.  For example, the oblique U parameter may or may not receive contributions from
operators affecting the  vector-boson vacuum polarization depending on the
basis, and yet the physical observable remains unchanged.~\cite{oblique.U}
Of course, depending on the renormalization group flow in the infrared, certain choices may be more appropriate than others, as emphasized in Ref.~\cite{georgi}.  In our examples, we generally eliminated redundant operators having more derivatives in favor of those having fewer.   To do otherwise would introduce inverse powers of possibly small masses into the effective couplings, defeating our desire to have an effective field theory describing the low-energy physics that is as insensitive as possible to infrared singularities.  It also conflicts with the extremely convenient mass-independent renormalization schemes in which, essentially, the mass-terms may be thought of as perturbations on the massless theory.

To summarize, so long as there is a consistent small parameter involved in making a perturbation expansion, one may use field redefinitions to justify application of the equivalence theorem.\cite{arzt}  Armed with the results in this paper, the analysis of running couplings associated with the reduced Lagrangian $\lredu$ is as economical as possible and properly accounts for the complications associated with counterterms for redundant operators.  In traditional applications, the infrared relevant and marginal operators are included in $\lr$ and infrared irrelevant operators in $\lh,$ but for strong coupling applications, for chiral perturbation theory, or in $1/N$-expansions, the groupings may be otherwise.  Thinking in terms of effective field theory frees us from traditional assignments for both phenomenological and theoretical purposes, and the preceding discussion can be generalized to all such cases. 

\vfill
\end{document}